**The wisdom of sages: nuclear physics education, knowledge-inquiry, and wisdom-inquiry**

A. Alan Cottey[a)]

*School of Chemistry, University of East Anglia, Norwich NR4 7TJ, UK*

In this paper I address the difference between knowledge-inquiry and wisdom-inquiry (concepts introduced by N. Maxwell) in nuclear physics education, specifically in senior-level textbooks for first-degree physics students. Following on from an earlier study of 57 such textbooks, I focus here on a remarkable use of literary quotations in one of them. The nuclear physics textbook *Particles and Nuclei: an Introduction to the Physical Concepts*, by B. Povh et al opens with a (German) quotation from *Max und Moritz* which has been rendered, in the celebrated translation by C. T. Brooks, as "Not alone to solve the double/ Rule of Three shall man take trouble;/ But must hear with pleasure Sages/ Teach the wisdom of the ages." What the student gets however is technical material followed abruptly at the very end by the advice (from The Book of Jeremiah) "And it shall be, when thou hast made an end of reading this book, that thou shalt bind a stone to it, and cast it into the midst of Euphrates". From a study of these and other quotations and other features of this book I infer a strong desire to express something important about wisdom, which is however even more powerfully suppressed by the ideology of knowledge-inquiry. At the end of this paper I discuss briefly *wisdom of the ages* and *wisdom for our age.*

## I. INTRODUCTION

In *The Shadow of the Bomb: a study of degree-level nuclear physics textbooks*[1] I presented a textual analysis of 57 nuclear physics textbooks for senior-level physics degree students. That work investigates how the textbooks deal with an aspect that is relevant and important but almost wholly avoided, namely nuclear weapons. The phrase 'Hamlet without the Prince' accurately evokes the nature of the narratives presented in these textbooks, a feature that has remained notably constant over the sixty year span of publication dates of the texts surveyed, apart from a small temporary shift in the late 1980s. In many of the textbooks one may, if willing to pay close attention to apparently tiny details, discern a subliminal tension between the desire to live up to the consensual norms of science education (which include clarity, directness, consistency, balance, focus, and generality) and an embarrassment about nuclear weapons (which leads to features which include vagueness, euphemism, and inattention).

One of those textbooks, *Particles and Nuclei: an Introduction to the Physical Concepts* by B. Povh, K. Rith, C. Scholz, and F. Zetsche[2] uses literary quotations in an unusual way, which I consider to be highly significant and is accordingly the subject of the present paper. My study of this led me to the conclusion that these quotations, whose theme is the need for thought extending beyond narrow attention to technical matters, comprise a significant illustration and vindication of the ideas of N Maxwell on knowledge-inquiry versus wisdom-inquiry.[3]

## II. KNOWLEDGE AND WISDOM

### A. Strongly emphasise the physical concepts

The Preface of *Particles and Nuclei: an Introduction to the Physical Concepts* states on page VII

> PARTICLES AND NUCLEI … conveys the fundamental knowledge in this area, which is required of a student majoring in physics. On traditional grounds these lectures, and therefore this book, strongly emphasise the physical concepts.

Phrases like *fundamental knowledge* and *emphasise the physical concepts* are commonly used in the introductory sections of nuclear physics textbooks and part of the generally accepted meaning is that applications are given little or no attention. As indicated, however, in *The Shadow of the Bomb*, especially in the section on Asymmetry, there usually is some attention to applications but it is highly selective (in a way which is generally covert). In *Particles and Nuclei* the emphasis in almost all of the book is indeed on physical concepts but an exception is made for one application - a brief digression, marked with a ■ symbol, on Reactors (page 276). In my opinion, this digression, one-third of a page, is too condensed to be useful to a student who did not already know what it said. This is significant, because the expositions in *Particles and Nuclei* are usually thorough and helpful to the student. These remarks about the Preface of *Particles and Nuclei* and about asymmetry provide a context for an attempt to understand the book's use of literary quotations.

**B. Wisdom of the ages**

Chapter 1 opens with an untranslated quotation (*Particles and Nuclei* is a translation into English of an earlier German edition) from the *4th Streich* of Wilhelm Busch's *Max und Moritz*

> Nicht allein in Rechnungssachen
>
> Soll der Mensch sich Muehe machen;
>
> Sondern auch der Weisheit Lehren
>
> Muss man mit Vergnuegen hoeren.

Busch's illustrated story of seven tricks played by the young rascals Max and Moritz was published in 1865 and is widely considered a classic of a genre which continues to this day, especially in comic-strip form. The charming doggerel of *Max und Moritz* was re-created in English by Charles T. Brooks in 1871. The German and English texts and Busch's dynamic illustrations can be found in a readily available Dover volume, edited by H. A. Klein.[4] The introduction to Max and Moritz's Fourth Trick, of which *Particles and Nuclei* quotes the seventh to the tenth lines, reads

An old saw runs somewhat so

Man must learn while here below. –

Not alone the A, B, C,

Raises man in dignity;

Not alone in reading, writing,

Reason finds a work inviting;

Not alone to solve the double

Rule of Three shall man take trouble;

But must hear with pleasure Sages

Teach the wisdom of the ages.

(I shall return to *wisdom* later. It is worth noting here that, although *of the ages* would not appear in a literal translation, the rendering does capture aptly the spirit of the whole of the Fourth Trick.)

## C. A sense of purpose and value

Almost all of *Particles and Nuclei* follows the claim to *strongly emphasise the physical concepts* made in the Preface but in the last two chapters there are some hints which seem to advocate going beyond mere calculation. Section 19.4, on *Particle Physics and Thermodynamics in the Early Universe*, opens with a quotation from John Maynard Smith

In all societies men have constructed myths about the origin of the universe and of man. The aim of these myths is to define man's place in nature, and thus give him a sense of purpose and value.

Yet nothing in this section or the rest of the book connects explicitly with purpose and value.

**D. Our modern struggles**

The final two chapters are 19 *Nuclear Thermodynamics* and 20 *Many-Body Systems in the Strong Interaction*. The book breaks out from its technical language, abruptly, in the very last paragraph of the main text (page 340) which reads

Our modern struggles to improve our understanding are fought on two fronts: physicists are testing whether the modern standard model of elementary particle physics is indeed fundamental or itself 'just' an effective theory, and are simultaneously trying to improve our understanding of the regularities of the complex systems of the strong interaction.

**E. Thou shalt bind a stone**

Then comes a final quotation

And it shall be, when thou hast made an end of reading this book, that thou shalt

bind a stone to it, and cast it into the midst of Euphrates:

*Jeremiah* 51.63

## III. WHAT ARE WE TO MAKE OF THESE QUOTATIONS?

How do they relate to the rest of the book? A unique meaning of a text, or indeed of any

element of discourse, is not to be found. Philip Pullman,[5] for example, expresses this

view eloquently. Prominently on the front page of his website he writes

As a passionate believer in the democracy of reading, I don't think it's the task of the author of a book to tell the reader what it means.

The meaning of a story emerges in the meeting between the words on the page and the thoughts in the reader's mind.

In addition, as I have emphasised in *The Shadow of the Bomb*, no book is simply the work of the authors. Many other people, editors, readers of drafts, proofreaders, marketing staff and designers contribute at the production stages. These producers make judgements (based on market research or on their life experience) about how the prospective readers will receive the book. Publishing proprietors and shareholders have a, usually indirect, influence on the choices made by the staff who are directly involved in the production. And after publication, reviewers and readers have their turn and the book can sink with little trace or be influential and remain in print for many years.

Obviously there is an element of humour in some of the quotations in *Particles and Nuclei* but this does not prevent us from trying, as well as we are able, to understand the book as a whole, considered as a social product and an element in nuclear discourse. Humour can indeed be of special significance as it generally is used to express things that one or more persons would be uncomfortable to express or hear directly.

Let us try to understand the significance of the inclusion of the quotation from Jeremiah. My own immediate reaction on reading this was that it could not possibly be conscious jocular self-deprecation. Rather, it connected with *conveys the fundamental knowledge … strongly emphasise the physical concepts* in the Preface. A student who had understood the book had no further need to refer back to it. This take on the potency of understanding the ideas of physics is common among experienced physicists.

I still believe that this may be a part of the meaning of the inclusion of the quotation but the violence of Jeremiah's advice suggests that there is something more - darker and almost certainly not the conscious intention of those responsible for the book. It is generally known, in the cultures that will read the book, in English or in German, that Jeremiah is dark. *A Jeremiah* is a dismal prophet, a denouncer of the times. *A jeremiad* (In German *Jeremiade*) is a doleful complaint or lamentation, a list of woes.

## A. So Jeremiah wrote in a book

To understand the quoted verse, we need more context[6] ...

> 60 So Jeremiah wrote in a book all the evil that should come upon Babylon, even all these words that are written against Babylon.
>
> 61 And Jeremiah said to Seraiah, When thou comest to Babylon, and shalt see, and shalt read all these words;
>
> 62 Then shalt thou say, O Lord, thou hast spoken against this place, to cut it off, that none shall remain in it, neither man nor beast, but that it shall be desolate for ever.
>
> 63 And it shall be, when thou hast made an end of reading this book, that thou shalt bind a stone to it, and cast it into the midst of Euphrates:
>
> 64 And thou shalt say, Thus shall Babylon sink, and shall not rise from the evil that I will bring upon her …

The meaning of the word 'evil' in this passage will be clearer to modern readers in the New International Version of the Bible,[7] which refers instead to 'disaster(s)'. It is however necessary to be aware that evil was the centre of Jeremiah's concerns (one can suspect that it was an obsession). The 'evil', or disasters, that the Lord will visit upon Babylon is closely connected with the evil ways which provoked this punishment. Further, Jeremiah was equally censorious of the evil behaviour of his own people.

Although most readers of Povh et al's book will not have a detailed or precise knowledge of the context quoted above, or of the wider backstory in The Book of Jeremiah, they may be expected to have at least a vague sense that Jeremiah was concerned about egregious evil and divine retribution. The question therefore will be in the reader's mind, even if subliminally, 'what is the evil that is being hinted at?'

**B. Motivation?**

What might drive a nuclear physics textbook which, according to the Preface, conveys fundamental knowledge, is traditional and strongly emphasises the physical concepts, to turn abruptly, at its very end, to a quotation which (obscurely) suggests evil and retribution? Nuclear weapons – the grotesque fission bomb, the much worse fusion bomb, the arsenals with many thousands of such bombs, their development at enormous expense of money and human skill, the rhetoric of threats, the fear, distrust and hatred – all of this could very reasonably be the object of a jeremiad.

Now one can see a connection with *the wisdom of the ages*, something those immature lads Max and Moritz failed to learn from their teacher. Instead they played one of their naughty tricks on him. What was this fourth trick? ...They filled his meerschaum pipe with gunpowder.

---------------------------------

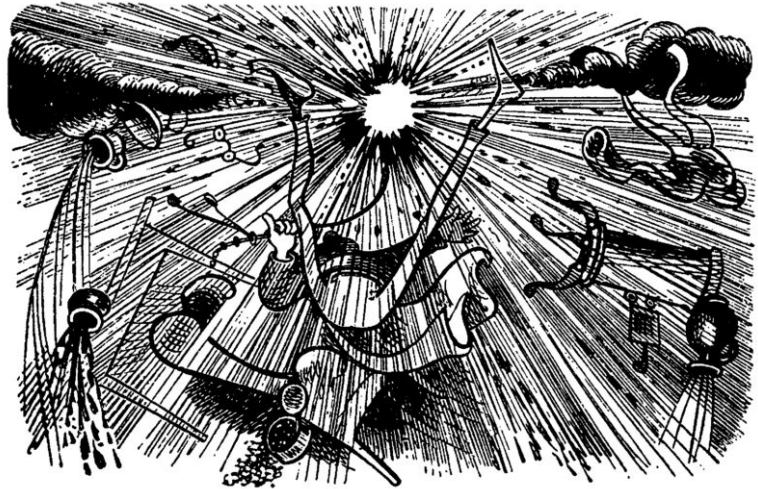

Fig. 1.[8] All are flying through the air/ In a lightning-powder-flash,[9]

<div align="center">--------------------------------</div>

Max and Moritz's infamous career passed through seven tricks but finally they got their comeuppance. They were put through the mill (literally) and fed to the ducks.

**C. Free from thermonuclear threats!**

In case any reader should still doubt a connection between this cautionary tale and nuclear affairs, I add finally that on the dedication page (iii) of the Dover edition of *Max and Moritz* the editor, H. Arthur Klein, writes

To the youngsters nearest now –

…

May their futures all be fine

& free from thermonuclear threats!

If this appears surprisingly direct, note that it was written in a time of justified anxiety about thermonuclear threats. The date of publication of the volume is 1962 and the Cuba missile crisis was in October of that year.

## D. Push of opposing forces

My interpretation of what is happening in *Particles and Nuclei* is that two incompatible needs are pushing against each other. One is a desire to express something about *wisdom of the ages* and *purpose and value*. The other is a commitment to a *traditional* style of physics education, which involves a focus on *fundamental knowledge*, strong emphasis of *physical concepts* and distancing from problematic social consequences. The traditional distancing can explain why the final brief paragraph *Our modern struggles … strong interaction* is tacked clumsily onto the end of much longer technical material.

Further, I suggest that the push of opposing forces lies behind the phrase *Our modern struggles to improve our understanding are fought.* This phrase betrays conflict between the pressures to suppress and to express something about the socially important consequences of nuclear physics.

**E. Empathy**

Since in my interpretation conflict is at the heart of *Particles and Nuclei* and indeed at the heart of my *Shadow of the Bomb* study, I add a few words about the approach to conflict – in any area, political, scientific or literary – advocated by Marshall Rosenberg[10] [11] which I find useful. His ideas rely heavily on *empathy* (defined by him on page 91 of *Nonviolent Communication* thus

> *a respectful understanding of what others are experiencing.*

For this study, empathy includes being constantly aware that none of us are *free from thermonuclear threats.* On the contrary we are overwhelmed by them. Countless evidences around us show that neither knowledge, nor education, prosperity, nationality, gender, or age comprises a remedy for the helplessness we feel.

The study of the 57 nuclear physics textbooks in *The Shadow of the Bomb* showed that remarkably little has changed since 1945. Further, what the nuclear physics books say and do not say and the manner in which they normalize the unnormalizable reflects practice in most branches of education and research. It seems that something is missing. What is it?

**IV. WISDOM FOR OUR AGE**

The study in this paper of the use of literary quotations in *Particles and Nuclei* indicates that nuclear weapons and their place in our culture are hard to address directly but also cannot be put entirely out of mind. We half believe but cannot fully admit that if, like the immature and foolish Max and Moritz, we do not take note of *the wisdom of the ages,* we will go through the mill and finish up as food for less foolish creatures.

If we take the story of Max and Moritz too seriously we will conclude that they are utterly deplorable and deserve their gruesome fate. I am more inclined to allow for Busch's ebullient energy and regard the story as a cheerfully grotesque exaggeration of the alienation which is common in boys. The world which the two lads experience with mischievous joy is created by adults who are generally conventional and unimaginative. Max and Moritz's teacher is evidently old-fashioned and pedantic and he has limited horizons. The perennial appeal of the terrible twosome's exploits surely has much to do with their iconoclasm.

In the real world, boys are often iconoclastic but they are redeemable. A question may now be asked about ourselves, supposedly adult and responsible ... are *we* redeemable? We hope so, of course, for we possess technologies and social structures capable of wreaking havoc far beyond the capabilities of little Max and Moritz.

What we need is *wisdom for our age* rather than *wisdom of the ages*. The latter, emphasising respect for authority and traditional values, was after all developed to deal with different conditions. In considering what would be the best understanding of the meaning of *wisdom*, I make use of the ideas of Nicholas Maxwell. He and Ronald Barnett describe it, on page viii of *Wisdom in the University*,[12] as

> the capacity to realize what is of value in life, for oneself and others, thus including knowledge and technological know-how, but much else besides.

*Wisdom-inquiry* is contrasted with *knowledge-inquiry*, meaning (page 4 of the article by Maxwell in the same book)

> Academic inquiry as it mostly exists at present, devoted to the growth of knowledge and technological know-how

And, again on page 4, the need for a change from a focus on knowledge to a focus on wisdom is explained thus

> Changing the nature of science, and of academic inquiry more generally, is the key intellectual and institutional change that we need to make in order to come to grips with our global problems – above all, the global problem behind all the others, the crisis of ever-increasing technological power in the absence of wisdom.

Maxwell has presented these and related ideas at greater length in *From Knowledge to Wisdom.*[13]

## V. CONCLUSION

Maxwell's take on knowledge and wisdom appears to me to be supported by the nature of *Particles and Nuclei*. I detect in the book a strong desire to express something important about wisdom, but also a powerful suppression of this desire by the ideology of knowledge-inquiry. I suggest that the negative attitude expressed in *The Book of the Prophet Jeremiah* is of limited help to us. The black humour in *Max and Moritz* and the constructive ideas expressed in *Wisdom in the University* have more potential.

**ACKNOWLEDGMENTS**

I acknowledge with thanks comments on drafts of this paper by T. Kibble, P. Le Mare, C. A. Mead, B. Panvel, M. Peters, P. Pickbourne, J. Satterthwaite and C. Wilson.

-------------------------------------[+]

a) Electronic mail: a.cottey@uea.ac.uk...

+